\documentstyle[11pt]{article}

\newcommand{\be}{\begin{equation}}
\newcommand{\ee}{\end{equation}}
\newcommand{\bea}{\begin{eqnarray}}
\newcommand{\eea}{\end{eqnarray}}
\newcommand{\ba}{\begin{array}}
\newcommand{\ea}{\end{array}}
\newcommand{\bc}{\begin{center}}
\newcommand{\ec}{\end{center}}
\newcommand{\bi}{\begin{itemize}}
\newcommand{\ei}{\end{itemize}}

\newcommand{\disregard}[1]{{}}

\def\bild#1\over#2{\mathrel{\mathop{\kern0pt #1}\limits_{#2}}}

\newcommand{\tr}[1]{\mathop{\mathrm{Tr}}\nolimits\left\{ #1 \right\}} 
\newcommand{\ket}[1]{|\kern.3ex#1\kern.3ex\rangle}
\newcommand{\bra}[1]{\langle\kern.3ex #1 \kern.3ex|}
\newcommand{\mean}[1]{\left\langle #1 \right\rangle} 
\newcommand{\sz}{\sigma_z}

\begin{document}

\centerline{\bf PERSISTENT CURRENTS and MAGNETIZATION }
\vskip .5cm
\centerline{\bf in TWO-DIMENSIONAL MAGNETIC QUANTUM SYSTEMS}
\vskip 1.5cm                 
\centerline{{\bf Jean DESBOIS},    
{\bf St\a'ephane OUVRY}
and {\bf Christophe TEXIER}}
\vskip 1cm
\centerline{{Division de Physique Th\'eorique \footnote{\it Unit\a'e de Recherche  des
Universit\a'es Paris 11 et Paris 6 associ\a'ee au CNRS},  IPN,
  Orsay Fr-91406}}

\begin{abstract} 
Persistent currents and magnetization  are considered for a two-dimensional
electron (or gas of electrons) coupled  to  various magnetic fields. 
Thermodynamic formulae for the magnetization and the persistent current are 
established and the
``classical'' relationship between current and magnetization is
shown to hold for  systems  invariant both by translation and rotation.
Applications are given, including the point vortex superposed
to an homogeneous magnetic field,
the quantum Hall geometry (an electric field and an homogeneous
magnetic field) and the random magnetic impurity problem (a random
distribution of point vortices).
\end{abstract}

\vskip 0.5cm


IPNO/TH 97-43

December 1997

\vfill\eject


\section{ Introduction }\label{sec:1}

\vskip1cm
   Since the pioneering work of Bloch \cite{bl}, several questions concerning
 persistent currents have been answered. 
 The
 conducting ring case has been largely discussed in the literature \cite{ri},
 as well as the persistent current due to  a point-like vortex \cite{vo}.
 
 However, many questions remain open. For example, 
  we might ask about standard relationships such as
   
 \be\label{n1}
 I= -\frac{dE}{d\Phi}
\ee
\be\label{n2}
M=IV
\ee
where $I,M,E,V,\Phi$ are, respectively, the persistent current,
 magnetization, energy, area (denoted by $V$ as volume in two dimensions) and 
 magnetic flux through the system.

These relations are clearly understood in the case of a ring.
 Do they apply  to more general, for instance infinite, systems? 
   More precisely, are they  still correct, and if yes, in which case?
   
   The aim of this paper is to clarify these points in the case of 
   two-dimensional systems. We will
   show that (\ref{n2}) holds true as soon as the system is invariant by
   translation and rotation. We will also insist on the role of the spin
   coupling, which happens to be crucial in the case of point vortices. 
   We will examplify these considerations for 
   an electron (or gas of electrons) coupled to
\vskip.3cm
-  a point vortex superposed to an homogeneous
magnetic field,
\vskip.3cm
- an uniform electric field and an homogeneous  magnetic field (quantum Hall
geometry),
\vskip.3cm    
-  a distribution of point-like 
  vortices randomly dropped onto the plane according to  Poisson's law,
  modelising  disordered magnetic systems (analytical and numerical).

The case of electrons on a ring threaded by a flux, and electrons in a plane
coupled to a point vortex will be revisited in  Appendix A.
Some technical details on harmonic regularization
for computing partition functions 
and the so called ``Landau counting rule'' will be given in Appendix B.
\vskip.5cm


\section{ Basic Definitions and Formalism}\label{sec:2}
\vskip.5cm
Consider the two-dimensional quantum mechanical problem of an
electron with Hamiltonian (we set the electron mass and the Planck constant
$m_e=\hbar=1$)
\be\label{n3}
H=\frac{1}{2}\left(\vec p -e\vec A \right)^2+V-\frac{e}{2}\sigma_zB
\ee
$V$ and $\vec A$ are, respectively, the  scalar and vector potential
 ($\vec B =\vec\nabla \times\vec A$) and $\sigma_z/2$ is the electron spin.
  The  local current is ($\vec v=\vec p-e\vec A$) 
\be
\vec j(\vec r)=\frac{e}{2}\left(\vec v\ket{\vec r}\bra{\vec r}+
\ket{\vec r}\bra{\vec r}\vec v\right)
\ee
The  total magnetization is  by definition
\be\label{n4}
M= \frac{1}{2}\int d\vec r\, (
   \vec r \times \langle{\vec j(\vec r)}\rangle)\cdot\vec k
   +  \frac{e}{2}\mean{\sigma_z} 
\ee
where $\vec k$ is the unit vector perpendicular to the plane and 
$\mean{\ \ }$ means average over Boltzmann
 or Fermi-Dirac distributions. In the Boltzmann case,
 one obtains the thermal magnetization ($Z_{\beta}\equiv \tr{e^{-\beta H}}$)
\be\label{n5}
M_{\beta}=\frac{e}{2Z_{\beta}}\tr{
 e^{-\beta H} \left( (\vec r\times \vec v )\cdot\vec k +\sigma_z  \right)
}
\ee

We want to relate the thermal magnetization  to a thermodynamical quantity.
To this aim,  we add a fictitious  uniform magnetic field $\vec B'$
perpendicular to the plane. In the symmetric gauge, its vector potential is
$e\vec A'=(e/2)\vec B'\times \vec r $.   $H$ becomes $H(B')$
  with the partition function $Z_{\beta}(B')\equiv \tr{e^{-\beta H(B')}}$.
First order
perturbation theory in $B'$ gives 
\be\label{n6}
Z_{\beta}(B')=Z_{\beta}+\frac{e\beta B'}{2}\tr{
e^{-\beta H}\left((\vec r\times\vec v)\cdot\vec k +\sigma_z\right)}+\ldots 
\ee
and thus,  the magnetization,
\be\label{n7}
M_{\beta}={1\over \beta}\lim_{B'\to 0}\frac{\partial 
 \ln Z_{\beta}(B')}{\partial B'}
\ee
In the sequel, we will  consider the spin-up and spin-down as two
distinct physical situations. It follows that in
(\ref {n7}), the spin induced part of the magnetization
is  $ {e\over 2} \sigma_z$, thus the orbital part of the magnetization is
\be \label{n71}
M_{\beta}^{orb}={1\over \beta}\lim_{B'\to 0}\frac{\partial 
 \ln Z_{\beta}(B')}{\partial B'} -{e\over 2}{\sigma_z}\ee

 Note  that in the particular case of $H$ containing  an homogeneous
   magnetic field $\vec B$ perpendicular to the plane, (\ref{n7}) narrows down
   to
   the usual formula
  \be\label{n70}
M_{\beta }=\frac{1}{\beta}\frac{\partial 
 \ln Z_{\beta}}{\partial B}
\ee
and (\ref{n71}) to
  \be\label{n70a}
M^{orb}_{\beta }=\frac{1}{\beta}\frac{\partial 
 \ln Z_{\beta}}{\partial B}-{e\over 2}{\sigma_z}
\ee
Note also that one can compute directly $M^{orb}_{\beta}$ using (\ref{n7})
by simply
 dropping the
coupling of the fictitious $\vec B'$ field to the spin in $Z_{\beta}(B')$,
and, accordingly, using (\ref{n70}) in the case of $H$ containing a
homgeneous $\vec B$ field, the coupling of $\vec B$ to the spin in
$Z_{\beta}$.

\vskip.3cm
Let us now turn to the persistent current and consider
 in the plane  a semi infinite line $\cal D$ starting at $\vec r_0$
 and making an angle
  $\theta_o$ with the horizontal $x$-axis. The orbital persistent
   current $I^{orb}(\vec r_0,\theta_0)$  through the line is
   \be\label{n8}
 I^{orb}(\vec r_0,\theta_0 )\equiv\int_{\cal D} d|\vec r-\vec r_0|\,
 {(\vec r-\vec r_0)\times \mean{\vec j(\vec r)}\over|\vec r-\vec r_0|}\cdot \vec k
 \ee
 where ${(\vec r-\vec r_0)\times {\vec j(\vec r)}\over |\vec r-\vec r_0|}\cdot \vec k
 $ is the
   orthoradial component of $\vec j(\vec r)$ to the line. It
 obviously depends on
   $\vec r_0$ and $\theta_0$.
   Consider now systems rotationnally invariant around $\vec r_0$.
 $I^{orb}(\vec r_0,\theta_0 )$ no longer depends on $\theta_0$ and, without loss
 of generality,  can be
 averaged over
 $\theta_0$. So
 \be\label{n9}
 I^{orb}(\vec r_0)=\frac{1}{2\pi }\int d\vec r
 {(\vec r-\vec r_0)\times \mean{j_{\theta}(\vec r)}\over 
 |\vec r-\vec r_0|^2}\cdot\vec k
  \ee
 and for the  Boltzmann distribution
 \be\label{n10}
 I^{orb}_{\beta}(\vec r_0)=\frac{e}{2\pi Z_{\beta}} \tr{
 e^{-\beta H}\frac{(\vec r -\vec r_0)\times \vec v}{\vert\vec
 r -\vec r_0\vert ^2}\cdot\vec k  }
 \ee
 Again, we want
 to establish a link between (\ref {n10}) and a thermodynamical quantity.
  We add to the system a fictitious  vortex $\phi'$ located at $\vec r_0$
  the origin of
  the line, with potential vector
  and magnetic field
\bea\label{n11}
e\vec A_v&=&\alpha '\frac{\vec k\times(\vec r-\vec r_0)}{|\vec r-\vec r_0|^2}
\\
e\vec B_v&=&2\pi \alpha'\delta (\vec r-\vec r_0)\vec k
\eea
  where $\alpha'=\phi'/\phi_o$ -$\phi_o$ is the flux quantum-  can always be chosen in the interval
 $[0,1]$ by periodicity.  The resulting Hamiltonian is
 \be \label{n711} H(\alpha ')= 
 \frac{1}{2}\left(\vec p -e\vec A -e\vec A_v\right)^2+V
 -\frac{e}{2}\sigma_z(B+B_v)
 \ee
 with partition function  $Z_{\beta }(\alpha ')$. 

 Though being a well-established fact
  \cite{stefan},
 we recall here that
 the spin coupling term $\sigma_z (B+B_v)$  in (\ref{n3},\ref{n711}) is
  necessary
 to  define  in a non ambiguous way point vortices quantum systems.
The behavior of the  wavefunctions at the location
$\vec r_0$ of the fictitious point vortex (or of a physical point vortex contained
in the original Hamiltonian (\ref{n3})) has to be specified
properly. This is  precisely achieved in a scale invariant way 
with the spin coupling term, which materializes  either
a repulsive hard-core (spin-down) $\pi\alpha'\delta(\vec r -\vec r_0)$ or attractive
(spin-up) $-  \pi\alpha'\delta(\vec r -\vec r_0)$ point contact interaction
 at $\vec r_0$.
In the standard  Aharonov-Bohm problem,
  the wavefunctions are chosen to vanish at $\vec r_0$, i.e. the spin-down
  repulsive prescription. However, we might  as
  well take
  the  attractive spin-up prescription.  
   Still, the persistent current   should
clearly
 be insensitive to the nature of the fictitious vortex introduced to define
 it,  and in particular on the way it is regularized at short
 distance. 

 First order perturbation theory in $\alpha'$ gives
 \be\label{n12}
 Z_{\beta }(\alpha ')=Z_{\beta } +\alpha '\beta \left(
 \tr{e^{-\beta H}\frac
 {(\vec r-\vec r_0)\times \vec v}
 {\vert \vec r-\vec r_0\vert ^2}\cdot\vec k} +\pi \sigma_zG_{\beta }(\vec
 r_0,\vec r_0)\right)+\ldots
 \ee
where
$G_{\beta }(\vec r,\vec r\,')\equiv \bra{\vec r}e^{-\beta H}\ket{\vec r\,'}$ is the thermal
propagator for the Hamiltonian (\ref {n3}).
We define the total current around $\vec r_0$ as 
\be\label{n120} I_{\beta}(\vec r_0)\equiv \frac{e}{2\pi\beta }\lim_{\alpha'\to 0}
\frac{\partial \ln Z_{\beta } (\alpha'  )}
{\partial \alpha' }\ee
and, from (\ref {n10}), 
\be \label{n13} 
I^{orb}_{\beta}(\vec r_0)=\frac{e}{2\pi\beta }\lim_{\alpha '\to 0}
\frac{\partial \ln Z_{\beta } (\alpha ' )}
{\partial \alpha ' }-{e\over 2} 
  \sigma_z{G_{\beta }(\vec r_0,\vec r_0)\over Z_{\beta}}
\ee
Note that in the particular case of
$H$ containing  a vortex $\phi$ at
$\vec r_0$, or consisting of a ring of center $\vec r_0$
threaded by a magnetic flux $\phi=\alpha\phi_0$, 
the orbital persistent current generated in the plane
around $\vec r_0$, or in the ring\footnote{In the case of a ring, 
the propagator at
the center of the ring is by definition nul.},
(\ref{n13}) narrows down to 
\be \label{n130}
I^{orb}_{\beta}(\vec r_0)=\frac{e}{2\pi\beta }
\frac{\partial \ln Z_{\beta } }
{\partial \alpha  } -{e\over 2} 
  \sigma_z{G_{\beta }(\vec r_0,\vec r_0)\over Z_{\beta}} 
\ee

If the first term in (\ref{n13})
is usually discussed
 in the literature under the weaker form appearing in  (\ref{n130}) valid for  the point
 vortex
  and the ring \cite{bl}-\cite{vo}, the second term in  (\ref{n13},\ref{n130}) triggered by the
 spin coupling to the propagator
 has been so far ignored. However,
  both terms are crucial, and  if the spin term is
  absent, some
 contradictions  do arise in the computation of orbital persistent currents.
 Indeed, orbital persistent currents -as well as  orbital magnetizations-
 should vanish when $H$ becomes free, and this is not the case if the spin
 term is ignored, 
 as we
 will see later for the point vortex, or for the 
 point vortex superposed to an
 homogeneous field. We will also encounter in the attractive  
 point vortex case a situation where the
 orbital current by itself  is not defined (in fact infinite), and where 
 only the total (orbital + spin) current has a
 sensible
 physical meaning. 

 To summarize, orbital magnetizations and currents are respectively given by
 (\ref{n71}) and (\ref{n13}). They should both vanish when the original 
 Hamiltonian $H$ becomes free.
 Total magnetizations and currents  are given by (\ref{n7}) and (\ref{n120}).
They obviously depend, as well as their orbital counterparts, 
on the spin degree of freedom defined in the original Hamiltonian (\ref{n3}).

If we consider systems that are both invariant by translation and
 rotation, we can proceed further by averaging (\ref {n10}) over $\vec r_0$. Taking advantage of
 the identity \cite{DFO}
  \be\label{n14}
 \int d\vec r_0\frac {(\vec r-\vec r_0)\times \vec v}{\vert \vec r-\vec
 r_0\vert ^2}=\pi\vec r\times\vec v
 \ee
 we can write, using (\ref {n5},\ref {n10}),
 \be\label{n15}
 I^{orb}=\frac{1}{V}\int d\vec r_0 I(\vec r_0)=\frac
 {1}{V}M^{orb}
 \ee
 which generalizes the conducting ring situation to 
 translation and rotation invariant two-dimensional  systems.
 Furthermore, since for a translation invariant system $G_{\beta }(\vec
r_0,\vec r_0) =Z_{\beta}/V$, (\ref{n15})
holds also for the total current and magnetization 
\be\label{n150} I={1\over V}M\ee
 Finally, (\ref{n15},\ref{n150})   still hold for non rotational invariant systems
 if the persistent current $I^{orb}$
   is understood as averaged over all directions $\theta_0$ of the semi infinite
    straight line.

To have access to the orbital  magnetization and persistent currents of a gas of
non-interacting electrons,
  $Z_{\beta}(B')$ or  $Z_{\beta}(\alpha')$ have to  be replaced in
  (\ref{n71},\ref{n13}), and $Z_{\beta}$ in 
(\ref{n70a},\ref{n130}),  by the corresponding grand-partition functions.   
For instance, (\ref{n70a}) and (\ref{n130}) respectively become   
\be
M^{orb}=\frac{1}{\beta}\frac{\ln\Xi_{\beta}(\mu)}{\partial B} - N
{e\over 2}\sz
\ee
\be
I^{orb}=\frac{e}{2\pi\beta}\frac{\partial}{\partial\alpha}\ln\Xi_{\beta}(\mu) -
\frac{e}{2}\sz\tr{\delta(\vec r-\vec r_0)f(H)}
\ee
where $\Xi_{\beta}(\mu)$ is the grand-partition function
-$\ln\Xi_{\beta}(\mu)=\tr{\ln(1+e^{-\beta(H-\mu)})}$- for chemical
potential $\mu$ and 
$f(E)$ is the Fermi-Dirac distribution. 
At zero temperature,
the second quantized magnetization and persistent current at
Fermi energy $E_F$ are by definition, using
the integral representation of the step function $\theta(E_F-H)$,
\be \label{mfermi}
M_{E_F}=
{1\over 2\pi i}\int_{-\infty}^{\infty}{dt'}{e^{iE_Ft'}\over
t'-i\eta'}Z_{\beta}M_{\beta}|_{\beta\to it'+\epsilon'}
\ee
\be \label{fermi}
I_{E_F}=
{1\over 2\pi i}\int_{-\infty}^{\infty}{dt'}{e^{iE_Ft'}\over
t'-i\eta'}Z_{\beta}I_{\beta}|_{\beta\to it'+\epsilon'}
\ee
where ${\eta',\epsilon'\to 0^+}$.
One should bear in mind that the number of electrons $N$
may be fixed, implicitly determining $E_F$ by
\be 
N= \frac{1}{2\pi i}\int_{-\infty}^{\infty}dt'\frac{e^{iE_Ft'}}{t'-i\eta'}
Z_{\beta}|_{\beta\to it'+\epsilon}
\ee

We insist here that we regard the first quantized thermalized magnetizations
and currents (\ref{n71},\ref{n13}), or (\ref{n7},\ref{n120}), as the
basic objects from which their second quantized versions can be
computed, for instance
via (\ref{mfermi}) and (\ref{fermi}) at zero temperature.
 In particular if $M_{\beta}$ or $I_{\beta}$ happens  not to depend on the
 temperature, that is to say if the magnetization or current for a quantum
 state do not depend on
 the energy of the quantum state, then the second quantized
 magnetization and current are simply given by $M=N M_{\beta}$ or
$I=N I_{\beta}$ - see for instance in Appendix A  the current for
the point vortex case.

\vskip.8cm


\section{Applications}\label{sec:3}
\vskip.5cm

\subsection{ The homogeneous magnetic field case}\label{subsec:3.2}
\vskip.3cm

As a warm-up exercise,  consider an homogeneous magnetic field
perpendicular to the plane. 
The spin-up ($\sz=+1$)
or down ($\sz=-1$)
partition function 
\be\label{n16} 
Z_{\beta}^{\sz}=\frac{V}{2\pi\beta}\frac{be^{\sz b}}{\sinh b}
\ee
where $b\equiv\beta\omega _c$ and $\omega_c=eB/2$
($eB>0$ is assumed without loss of generality) leads to the orbital magnetization 
\be\label{n17}
M_{\beta}^{orb}={1\over \beta}{\partial \ln Z^{\sz}\over \partial B}
-\sz{e\over 2}
=\frac{e}{2}\left(\frac{1}{b}-\coth b \right)
\ee
which in this particular case does not depend on $\sz$.

Since the system is invariant by translation and rotation,
we know a priori that  the orbital persistent current is given by
$I^{orb}_{\beta}=M_{\beta}^{orb}/V$.
Let us however illustrate the considerations above by computing
 directly the orbital persistent current around a point $\vec r_0$.
Adding a vortex $\alpha'$ located at this point with a repulsive
 (spin-down) or attractive (spin-up) prescription, we get \cite{chris}
 for $0\le\alpha'\le 1/2$  the partition function (see Appendix B)
\be\label{n18}
Z_{\beta}^{\sz}(\alpha')=Z_{\beta}^{\sz}+\frac{e^{\sz b}}{2\sinh b}
\left(
  \alpha' - e^{-(\sz+\alpha')b}\frac{\sinh\alpha' b}{\sinh b}
\right)
\ee
One has also
\be
 G_{\beta}^{\sz}(\vec r_0,\vec r_0)=\frac{1}{V} Z_{\beta}^{\sz}
\ee
Thus, (\ref{n13}) gives  the orbital persistent current 
\be\label{n19}
I^{orb}_{\beta}=\frac{e}{2\pi\beta }\lim _{\alpha'\to 0}\frac{\partial \ln
Z_{\beta}^{\sz}(\alpha')}{\partial\alpha' }
-\sz \frac{e}{2V}=\frac{e}{2V}\left( \frac{1}{b}-\coth b
\right)
\ee
which does not depend on $\vec r_0$, and indeed satisfies
$M_{\beta}^{orb}=I_{\beta}^{orb}V$.
 Note that it does not depend on $\sz$, i.e. in particular on the short
 distance behavior of the
fictitious vortex.
  \vskip.3cm

\subsection{ The point vortex  + a homogeneous magnetic field case}\label{subsec:3.3}
\vskip.3cm

We consider a point vortex carrying a  flux $\phi/\phi_0=\alpha$
at the origin of the plane superposed to a homogeneous $\vec B$
field, and concentrate on the repulsive spin-down case, bearing in mind
that the
analysis in the
attractive  spin-up case would follow the same lines as in the point vortex case (see
Appendix A).
Using
the partition function (\ref{n18}) with $\alpha'\to\alpha$, $\sz=-1$, 
(\ref{n70a}) gives 
 the orbital magnetization in the spin-down case
\bea\label{n20}
M_{\beta}^{orb}&=&\frac{e}{2}(\frac{1}{b}-\coth b)
-\frac{e\pi\beta}{2V}
\frac{1}{b^2}\bigg(
      \alpha-e^{-(\alpha-1)b}\frac{\sinh\alpha b}{\sinh b}
  \nonumber\\& &\hspace{3cm}
  +\frac{b}{\sinh b}e^{-(2\alpha-1)b}(
    \alpha-e^{(\alpha-1)b}\frac{\sinh\alpha b}{\sinh b}
  )
\bigg)+\ldots 
\eea
where  $\ldots$ denote corrections which are subleading in volume.
The vortex does not affect the leading volume term, which
is therefore identical to the pure homogeneous magnetic field magnetization
(\ref {n17}).

Let us now turn to the persistent current around the vortex location at the
origin of the plane. 

Since, when $\alpha> 0$,
\be
G_{\beta}(\vec 0, \vec 0)=0
\ee
$I_{\beta}(\vec 0)=I^{orb}_{\beta}(\vec 0)$. From
(\ref{n130}), 
the persistent current at leading order in volume reads
\be\label{n21}
I^{orb}_{\beta}(\vec 0)=\frac{e}{2V}\left(\frac{1}{b}-\frac{2e^{-2\alpha b}}
 {1- e^{-2b}} \right)
\ee
with indeed $M^{orb}_{\beta}\ne I^{orb}_{\beta} V$.

When $\alpha=0$ on the other hand, the orbital persistent current should narrow down
to the homogeneous magnetic field result (\ref{n19}). 
However, this does not happen for the current in (\ref{n21}).
But, if one pays attention to the fact that, when $\alpha=0$,
\be 
G_{\beta}(\vec 0, \vec 0)= \frac{Z_{\beta}}{ V}
\ee
where $Z_{\beta}$ stands for the partition function of the homogeneous
$\vec B$ field,
then the spin term is such that in (\ref{n130}) the orbital persistent current correctly
narrows down to 
the orbital current for the homogeneous magnetic field, i.e. 
\be 
I^{orb}_{\beta}=\frac{e}{2V}\left(\frac{1}{b}-\frac{2}{1- e^{-2b}} \right)
+{e\over 2V}={e\over 2V}({1\over b}-\coth b)
\ee

\subsection{ The case of the plane with an electric field }\label{subsec:3.4}

 Consider now a   homogeneous magnetic field
  and an  electric field $\vec E$ in the horizontal
 $x$-direction (Hall geometry) and
ask how the magnetization and the persistent current (\ref{n17},\ref{n19}) are
 affected by the electric  field.
The spectrum  is
unbounded from below, making the use of thermodynamical quantities as the
partition function hazardous. We propose to circumvent this
difficulty by confining the electron in  a harmonic well (for more details
on harmonic regularization see Appendix B). The
Hamiltonian in the symmetric gauge $\vec A=B\vec k\times\vec r/2$ reads
\be H={1\over 2}(\vec p-e\vec A)^2-eEx+{1\over 2}\omega^2\vec r^2\ee
with indeed a bounded spectrum.
Its propagator is
\be G_{\beta}(\vec r, \vec r\,')= G_{\beta}^{(B,\omega)}(\vec r-e{\vec E\over
\omega^2},\vec r\,'-e{\vec E\over \omega^2})
e^{-i\omega_c{e\vec E\over \omega^2}\cdot\vec k\times (\vec r-\vec r\,')}e^{{\beta\over 2}{({eE\over
\omega})^2}}
\ee
where
\be
G_{\beta}^{(B,\omega)}(\vec R,\vec R\,')=
{\omega_t \over 2\pi\sinh\beta\omega_t}
e^{
  -{\omega_t\over 2\sinh\beta\omega_t}
  \big(
  \cosh\beta\omega_c(\vec R-\vec R\,')^2+2i\sinh\beta\omega_c
  (\vec R\times\vec R\,')\cdot \vec k
  +(\cosh\beta\omega_t-\cosh\beta\omega_c)(\vec R^2+\vec R\,'^2)
 \big)
}
\ee
is the propagator for
the homogeneous $\vec B$ field
in a  harmonic well
and $\omega_t=\sqrt{\omega^2+\omega_c^2}$.
Note that in the thermodynamic limit  $\omega\to 0$, 
\bea\label{coin}  
G_{\beta}(\vec r, \vec r\,')={\omega_c\over
2\pi\sinh\beta\omega_c}
\exp\Big\{-{\omega_c\over
2\sinh\beta\omega_c}[\cosh\beta\omega_c(\vec r-\vec r\,')^2
 +2i\sinh\beta\omega_c(\vec r\times\vec
r\,')\cdot\vec k\nonumber\\
-{\sinh\beta\omega_c}(x+x'){\beta eE\over \omega_c}
+{\sinh\beta\omega_c\over 2}({1\over \beta\omega_c}-\coth\beta\omega_c)
{\beta eE\over \omega_c}({\beta eE\over
\omega_c}+2i(y-y'))]
\Big\}
\eea
 still depends on $\vec r$ at coinciding points  $\vec r=\vec r\,'$,
a direct manifestation of the breaking of
translation invariance.
The partition function  reads (without the spin coupling to the $\vec B$
field since we are interested in the orbital magnetization only)
\be Z_{\beta}= {1\over 2}{1\over \cosh\beta\omega_t-\cosh\beta\omega_c}e^{{\beta\over
2}({eE\over\omega})^2}\ee
If
$E\ne 0$, the
partition function
blows up in the thermodynamic limit
$\omega\to 0$.
On the contrary, when $E=0$, it correctly yields back
the partition function for the homogeneous
magnetic field
 if the proper thermodynamic limit prescription
$\lim_{\omega\to 0}{{2\pi\beta}\over (\beta\omega)^2}\to V$ is taken. 

The orbital magnetization reads
\be\label{magE} M_{\beta}^{orb}={e\over
2}{\sinh\beta\omega_c-{\omega_c\over\omega_t}\sinh\beta\omega_t\over
\cosh\beta\omega_t-\cosh\beta\omega_c}\ee
which happens to coincide with (\ref{n17}) in the thermodynamic limit.
 The electric field has  no effect on the magnetization of the system, a
 result that can be understood at the semi-classical level by the fact that
 the electric field
  only induces a translation of cyclotron orbits.

Turning now
to the persistent current around a point $\vec r_0$, a fictitious vortex $\alpha'$
located at $\vec r_0$ is added to the system, that we take to be
repulsive. 
All what is needed is the partition function of an homogeneous magnetic field
in presence of an electric field and a harmonic well,
and  a repulsive vortex at $\vec r_0$, at
leading order in $\alpha'$. It can be obtained in perturbation theory 
along
the lines developed in \cite{deveigy}.
The harmonic well has not only the virtue, as we have seen,
to make tractable the
computation of an otherwise diverging partition function with an $\vec E$ field, 
but it is also needed \cite{deveigy}
to give an unambiguous meaning to the
perturbative analysis in $\alpha'$, which is otherwise ill-defined.

The first order perturbative correction  to the partition function 
$Z_{\beta}(\alpha')=Z_{\beta}+\alpha' Z_{\beta}^1+\ldots $
\be 
Z^1_{\beta} = 2\beta\int dzd{\bar z}  {1\over
\bar z-\bar z_0}(\partial_z-{1\over 2}\omega_c \bar z)
G_{\beta}(\vec r,\vec r\,')\big|_{\vec r\,'=\vec r}
\ee
where $z=x+iy$ is the complex coordinate in the plane, gives an orbital  persistent current
\be I^{orb}_{\beta}(\vec r_0)=    {e\over 2\pi\beta}
{Z_{\beta}^1\over Z_{\beta}}+ {e\over 2Z_{\beta}}G_{\beta}(\vec r_0,\vec
r_0)\ee
equal to
\be\label{E} I^{orb}_{\beta}(\vec r_0)=    
{e\over 2\pi} e^{-{\omega_t\over
\sinh\beta\omega_t}(\cosh\beta\omega_t-\cosh\beta\omega_c)|\vec r_o-e{\vec
E\over \omega^2}|^2}
{\omega_t\over \sinh\beta\omega_t}
\left(\sinh\beta\omega_c-{\omega_c\over\omega_t}\sinh\beta\omega_t\right)\ee
The current is exponentially damped when the point $\vec r_0$ around
which it is estimated differs from the equilibrium position in the  harmonic
well $\vec r_o-e{\vec
E\over \omega^2}=\vec 0$.
When $\vec E=\vec 0$, (\ref{E}) coincides, in the thermodynamic
limit, with the orbital current
for the homogeneous magnetic field (\ref{n19}), as it should.  When $\vec E\ne 0$, 
this is still the case 
if $\vec r_0=e\vec E/\omega^2$, which means that the effect of the electric
field it to locate the current at
the edge of the sample. However, 
as soon as  $\vec r_0\ne e\vec
E/\omega^2$, the orbital curent 
$I^{orb}_{\beta}(\vec r_0)$ vanishes in the thermodynamic limit.


\section{The Magnetic
 Impurity Problem}\label{sec:4}

\vskip.4cm
Consider  \cite{DFO} a random Poissonian distribution
 $\{ \vec r_i, i=1,2,\ldots ,N\}$ of $N$ hard-core vortices at position $\vec r_i$,
  carrying a flux $\phi=\alpha \phi_o$, where $\alpha$ can always be taken
  in the interval $[0,1/2]$ without loss of generality. After averaging over disorder, 
  the system is
  invariant by
    translation and rotation, so it is sufficient to compute $M^{orb}$ to get
     the orbital persistent current, since they are  related by
     (\ref{n2}). The Hamiltonian is 
\be
  H=\frac{1}{2}\left(\vec p-e\vec A(\vec r) \right)^2+\frac{e}{2}B(\vec r)
\ee
with 
\be\label{n24}
e\vec A(\vec r)=\alpha\sum_{i=1}^N\frac{\vec k\times (\vec r-\vec r_i)}
{\vert\vec r-\vec r_i\vert^2}
\ee
\be
eB(\vec r)=2\pi\alpha\sum_{i=1}^N\delta(\vec r-\vec r_i) 
\ee
$B(\vec r)$ might be replaced by its mean value 
\be\label{n25}
\mean{B}=\frac{2\pi\rho\alpha }{e}
\ee
a mean field approximation that is valid in the limit
 $\alpha\to 0$, and $\rho\equiv N/V$, the vortex density, fixed \cite{DFO}.

\vskip.3cm
In a Brownian motion approach
 \cite{DFO}, for a given configuration of vortices, 
 the partition   function is
\be\label{n26}
Z_{\beta }=Z_{\beta }^{o}\left< \exp\left(i\sum_{i=1}^N 2\pi n_i\alpha
\right) \right>_{\{C\}}
\ee
where $\mean{\ \ }_{\{C\}}$ means averaging over all  closed Brownian curves
 of length $\beta$, and  $n_i$ is the winding number around the vortex at
 location $\vec r_i$. $Z_{\beta}^o=V/(2\pi\beta)$ is the free partition
 function. 

Averaging over all vortices configurations leads to  
\be\label{n27}
Z_{\beta }=Z_{\beta}^o\left< \exp\left(
\rho\sum_nS_n(e^{2i\pi\alpha n}-1)\right) \right>_{\{C\}}
\ee
where $S_n$ stands for the arithmetic area of the $n$-winding sector of the
 given Brownian curve chosen in $\{C\}$. Due
 to  scaling properties of Brownian curves
 \cite{DFO,cdo},
 the random variable $S_n$ scales like $\beta$, 
  thus  the rescaled variables  
\be\label{n28}
S=\frac{2}{\beta }\sum_nS_n\sin ^2(\pi\alpha n)\quad \quad
 A=\frac{1}{\beta }\sum_nS_n\sin (2\pi\alpha n)
\ee
The probability distributions of $S$ and $A$ are actually independent of
 $\beta$ with 
  $\mean{S}_{\{C\}}=\pi\alpha (1-\alpha )$ and  $\mean{A}_{\{C\}}=0$.
 
\vskip.2cm
To get the magnetization, we add a fictitious magnetic field $\vec B'$, with
partition function (without the spin  coupling to the fictitious $\vec B'$
field since we are interested in the orbital magnetization only)
 
\be\label{n30}
Z_{\beta }(B')=Z_{\beta}^o\left<e^{-\beta\rho (S-iA)+i\beta B'\cal A}
\right>_{\{C\}}
\ee
where ${\cal {A} }\equiv (1/\beta )\sum_n nS_n $ is the rescaled algebraic area
 enclosed by the Brownian curve in $\{C\}$. From (\ref{n7}), and
since $S\to S$, $A\to -A$, and $\cal A\to -\cal A$ when one spans a 
Brownian curve  in the opposite direction, the orbital magnetization is
\be\label{n31}
M_{\beta}^{orb}=
-\frac{\left<  e^{-\beta\rho S}\sin (\beta\rho A)\cal A \right>_{\{C\}}}
{\left< e^{-\beta\rho S}\cos (\beta\rho A) \right>_{\{C\}}}
\ee
$M_{\beta}^{orb}$ is  actually only a function of
 $\beta\rho$ and $\alpha$, odd in $\alpha$.
 Thus, for $\alpha\in [0,1/2]$, necessarily
 \be\label{n32}
M_{\beta}^{orb}=(1-2\alpha )F(\beta\rho ,\alpha (1-\alpha ))=
(1-2\alpha )\sum_{n=1}^{\infty }(\beta\rho )^n\sum_{m\ge n}a_{mn}
(\alpha(1-\alpha ))^m
\ee
which can in principle be obtained in perturbation
 theory \cite{DFO,chris}. 
 
 The $\rho ^n\alpha ^m$ term describes
 an electron interacting $m$ times with $n$ vortices, so $m\ge n$.
   The mean magnetic field
  (\ref{n25}) scales like $\rho\alpha$, therefore the mean
  field terms are those with $m=n$. They are in fact  easy to obtain,
since  $\alpha\to 0$ corresponds to the mean field regime. Thus, from
(\ref{n17},\ref{n25},\ref{n32}), 
\be\label{n33}
{\mean M_{\beta}^{orb}}=\frac{e}{2}\left({1\over {\mean b}}
- \coth {\mean b}\right)
\ee
where ${\mean b}=\beta\mean{\omega_c}$, with
$\mean{\omega_c}=e\mean{B}/2=\pi\rho\alpha$.
From (\ref{n32}),  one infers that $M^{orb}_{\beta}$ necessarily contains
\be\label{n331}
M_{\beta}^{orb}|_{mean}=(1-2\alpha)\frac{e}{2}\left({1\over {\mean b}'}
- \coth {\mean b}'\right)
\ee
where ${\mean b}'=\beta\pi\rho\alpha(1-\alpha)$.

It is even possible to go a little bit further, by observing that the
 $\rho ^n\alpha ^{n+1}$ terms come from the homogeneous $\mean{B}$ field +
  one vortex case \cite{chris}.  
  The first correction to   the mean field   approximation (\ref{n331}) is
 \be\label{n34}
M_{\beta}^{orb}=M_{\beta}^{orb}|_{mean}+e\alpha (1-\alpha )(1-2\alpha)\left(
{1\over 2{\mean b}'}+\frac{{\mean b}'-1-e^{-2{\mean b}'}}{1-e^{-2{\mean b}'}}
 +\frac{2{\mean b}'(1-{\mean b}')e^{-2{\mean b}'}}{(1-e^{-2{\mean b}'})^2} \right)+\cdots 
\ee

\vskip.2cm
In Fig. 1 ($\beta\rho =1)$,  the agreement is rather good
between (\ref {n34}) - full curve - and the numerical simulations - points -
based on (\ref {n31}). We generated 2000 random walks of 100000 steps
each one. However, the situation becomes less transparent
for higher $\beta\rho$
values. Clearly, the perturbative analytical approach  need more and more
corrections coming from $\mean{B}$ + two vortices, \ldots.
The numerical simulations  indicate that, for $0<\alpha<1/2$,
$M_{\beta}^{orb}$ is  negative whatever $\beta\rho$ is. Even if it is physically clear why  it should be so since the
vortices enforce a clockwise rotation of  the electrons when 
$0<\alpha<1/2$, an analytical proof
is  missing.

\begin{appendix}

\section{Appendix A: Persistent current for the ring and the point vortex}\label{sec:A}

\subsection{The case of the ring revisited}

Consider a ring of radius $R$ threaded by a homogeneous magnetic field $B$
perpendicular to the ring, with flux  $\phi$   through the
ring. Let us denote
as usual $\alpha=e\phi/2\pi=\phi/\phi_o$ the value of $\phi$ in unit of the quantum of
flux. The system is  periodic in $\alpha$ with period 1. Reversing the sign
of $B$ does not change the physics either -the partition function is
symmetric  around $\alpha=1/2$, whereas the magnetization and   current are
antisymmytric-, so one can always restrict
$0\le\alpha\le 1/2$. Clearly, the orbital persistent
current  changes its sign
with $\phi$.
But $\phi=\pm 1/2$ are identical, thus   necessarily
$I^{orb}_{\beta}(\alpha=1/2)=0$. 

To summarize, both  first and second quantized  orbital
currents should be defined  in the interval $\alpha\in [0,1/2]$, and
due to
obvious symmetry considerations, they should vanish when $\alpha=0, 1/2$.

In unit $2R^2=1$ the  partition function is
\be 
Z_{\beta}(\alpha)=\sum_{m=-\infty}^{m=\infty} e^{-\beta{(m-\alpha)}^2}
\ee

The thermal current is formally defined as in (\ref{n13}). However, it is
equivalent to
thread the ring with an homogeneous $\vec B$ field or at its center with a
vortex line  of flux $\phi$. It follows that
(\ref{n13}) narrows down to (\ref{n130}). Clearly no distinction has to be
 made either between the orbital and total magnetization (or current) since
 by geometry the
 propagator at the center of the ring is zero. Thus 
\be I_{\beta}=
I^{orb}_{\beta}= {e\over 2\pi\beta}{\partial\ln Z_{\beta}\over\partial\alpha}
\ee
and
\be M_{\beta}=
M^{orb}_{\beta}= {1\over\beta}{\partial\ln Z_{\beta}\over\partial B}
\ee
Since $\phi=\alpha 2\pi/e=B\pi R^2$, one  necessarily has
$M_{\beta}=
I_{\beta} \pi R^2$.

One obtains 
\be\label{i} I_{\beta}=-{e\over 2\pi}\left(2\alpha+2\sum_{k=1}^{\infty}
{\sinh(2k\beta\alpha)\over \sinh(k\beta)}(-1)^k\right)
\ee
or equivalently, using the reciprocity formula,
\be 
Z_{\beta}(\alpha)=\sqrt{\pi\over\beta}e^{-\alpha^2\beta}
Z_{\pi^2\over\beta}({i\alpha\beta\over\pi})\ee

\be\label{ii} I_{\beta}={e\over \beta}\sum_{k=1}^{\infty}
{\sin(2\pi k\alpha)\over \sinh(k\pi^2/\beta)}(-1)^k\ee
Clearly, from (\ref{i},\ref{ii}) 
\be I_{\beta}(\alpha=0)=0=
I_{\beta}(\alpha=1/2) \ee
as it should. Note that (\ref{i},\ref{ii}) are, to our knowledge, the simplest
expressions obtained so far for the thermalized persistent current on a ring.
(\ref{ii}) is the sine Fourier series of a periodic and odd function,
making the symmetries of $I_{\beta}$ explicit.

Considering now  a second quantized   gas of electrons
on the ring, starting from $I_{\beta}$,
the second quantized persistent current at zero temperature 
and Fermi energy $E_F$ is  given by (\ref{fermi}). 
However, since individual persistent currents for each quantum states are
somple to obtain,
one may  sum individual currents up to the Fermi energy\footnote{From
now on we revisite the approach followed in \cite{ri}.}.
The current carried by the state $|m>$ of orbital momentum $m$ and energy
$(m-\alpha)^2$ is $I_m={e\over \pi}(m-\alpha)$. It follows that
if $N$ is even,
\be\label{even} 
I_{E_F}={e\over 2\pi}N(1-2\alpha)
\ee
and if $N$ is odd,
\be\label{odd} 
I_{E_F}=-{e\over \pi}\alpha N
\ee
These results seem misleading since the currents in (\ref{even},\ref{odd}) do not vanish either
at $\alpha=0$ or  $\alpha=1/2$. However,
 their derivation in terms of individual quantum states
is valid for a spectrum with no degenerate
states, i.e. if and only if  $0<\alpha<1/2$.
Indeed, when $\alpha=0$, the states $|m>$ and $|-m>$ are degenerate, whereas
when
$\alpha=1/2$, the states $|1-m>$ and $|m>$ are degenerate.
It follows that (\ref{even}) is uncorrect
when  $\alpha=0$, and (\ref{odd})  is uncorrect when
$\alpha=1/2$.
If the degeneracy
is properly taken into account, meaning that a given  degenerate
doublet of states contributes to the current as
${I_m+I_{-m}\over 2}$ when $\alpha=0$, or
${I_{1-m}+I_m\over 2}$ when $\alpha=1/2$, the total current ends up  to
vanish as it should when $\alpha=0,1/2$.

We conclude that the persistent current for a gas of
electrons at zero temperature
on a ring threaded
by a flux $\alpha\phi_o$ is respectively given by (\ref{even}) or (\ref{odd})
when $N$ is even or odd
if $0<\alpha<1/2$, and $0$ if $\alpha=0,1/2$.

\subsection{ The case of point vortex revisited}

Consider finally a  point  vortex of flux $\phi=\alpha\phi_0$
at the origin of the plane, which can
be either hard-core (spin-down repulsive prescription), or attractive
(spin-up attractive prescription), clearly two different physical
situations.

As in the magnetic impurity and ring cases, one can always restrict the
interval of definition of $\alpha\in [0,1/2]$. However, to illustrate the
non trivial role of the spin, we will explicit the results in the whole
interval $\alpha\in[0,1[$.

The spin-down and (up) partition function respectively read
(see
Appendix B)
\be \label{hehe}
Z^{-}_{\beta}-Z_{\beta}^o={1\over 2}\alpha(\alpha-1) 
\hspace{1cm} 0\le\alpha < 1
\ee
\bea
Z^{+}_{\beta}-Z_{\beta}^o&=&{1\over 2}\alpha(\alpha+1) 
  \hspace{2cm} 0\le\alpha\le\frac{1}{2}\nonumber\\
Z^{+}_{\beta}-Z_{\beta}^o&=&{1\over 2}(\alpha-1)(\alpha-2) 
  \hspace{1cm}  {1\over 2}\le\alpha< 1  
\eea
leading to the total current around the origin\footnote{Computing the
current around a point $\vec r_0$ would require the
partition function of a vortex $\alpha$
at the origin and another fictitious vortex $\alpha'$ at $\vec r_0$
at leading order in $\alpha'$.}
\be\label{down} 
I^{-}_{\beta} =\frac{e}{V}\left( \alpha-{1\over 2} \right)
                          \hspace{1cm} 0\le \alpha<1
\ee
\bea\label{up} 
I^{+}_{\beta}&=&\frac{e}{V}\left( \alpha+{1\over 2} \right)
                          \hspace{1cm}  0\le\alpha< {1\over 2} 
\nonumber\\\label{upp}
I^{+}_{\beta}&=&\frac{e}{V}\left( \alpha-{3\over 2} \right)
                          \hspace{1cm} {1\over 2}<\alpha< 1
\eea
which is discontinuous at $\alpha=1/2$ in the spin-up case.

We would like to identify  the orbital part $I^{orb}_{\beta}$ of the currents
(\ref{down},\ref{up}).

If
$\alpha=0$, i.e. the free case, by definition
\be G^{+}_{\beta}(\vec 0,\vec 0)=  G^{-}_{\beta}(\vec 0,\vec
0)={Z_{\beta}^o\over V}\ee
One rightly deduces
\be 
I_{\beta}^{orb}=\lim_{\alpha\to 0} {e\over V}(\alpha+\sz{1\over 2})
-\sz{e\over 2V}=0
\ee
as it should.

If $0<\alpha<1$, 
in the repulsive spin-down case, 
\be G^{-}_{\beta}(\vec 0,\vec 0)= 0 \ee
We conclude that the orbital persistent current generated by a repulsive vortex at the
origin is
\be I^{orb-}_{\beta}=0 \quad \alpha=0\ee
\be I^{orb-}_{\beta}={e\over V}(\alpha-{1\over 2}) \quad 0<\alpha<1\ee

On the other hand,
in the attractive spin-up case, the propagator at
the location of the vortex is infinite as soon as $\alpha\ne 0$,
thus $I_{\beta}^{orb+}$  diverges. We conclude that the orbital persistent
current generated by an attractive vortex is not defined (in fact infinite), 
only the total
current (orbital + spin) is defined
and given by (\ref{up}).

The thermalized persistent currents (\ref{down},\ref{up}) do not depend
on the temperature.
It follows trivially that for a gas of $N$  electrons, in the
interval $0<\alpha<1/2$ for instance, the persistent current rewrites as
\be\label{current} 
I^{\sz}=e{\rho_e}(\alpha +{\sz\over 2})
\ee
where $\rho_e=N/V$ is the electron density.

The magnetization induced by a vortex in the plane (\ref{n70}) is deduced from
the partition function of a vortex  $\alpha$ + a
fictitious $\vec B'$ field,
i.e. (\ref{n18}), with  $\alpha'\to\alpha$ and $ \vec B\to \vec B'$.
One gets 
\be \label{n22}
M^{orb-}_{\beta}=-\frac{1}{V}\frac{e\pi\beta }{3}\alpha (1-\alpha)
(\frac{1}{2}-\alpha) \hspace{1cm} 0<\alpha<1
\ee
and 
\bea \label{n221}
M^{orb+}_{\beta}&=&-\frac{1}{V}\frac{e\pi\beta }{3}\alpha (1+\alpha)
(\frac{1}{2}+\alpha) \hspace{1.7cm} 0<\alpha<\frac{1}{2}\nonumber\\
M^{orb+}_{\beta}&=&\frac{1}{V}\frac{e\pi\beta }{3}(1-\alpha)(2-\alpha)
(\frac{3}{2}-\alpha) \hspace{1cm} \frac{1}{2}<\alpha<1
\eea
which are subleading in volume compared to the homogeneous field case.
The magnetization for a gas of electrons is, for
$0\le\alpha<1/2$,
\be 
M^{orb, \sz}=-\frac{e}{6}\alpha (\sz+\alpha)(\frac{\sz}{2}+\alpha)
\frac{1}{e^{-\beta\mu}+1}
\ee
with the additional Fermi factor at zero energy.
If it is to be compared to the current $I^{\sz}$ in (\ref{current}), one sees
that both 
the current and the magnetization induced by a point vortex in the plane
for a gas of electrons have the same leading volume behavior.


\section{Appendix B: Harmonic regularization and Naive Landau counting rule}
\subsection{Harmonic regularization}

Partition functions for systems with, in the
thermodynamic limit, a continuous
spectrum (point vortex) or a discrete but infinitely degenerated spectrum
(homogeneous magnetic field) can be obtained in a non ambiguous way
by regularizing the system at long distance by
adding an harmonic term $\omega^2\vec r^2/2$  to the Hamiltonian, computing
the
partition function in the presence of the regulator, and then taking the
thermodynamic limit $\omega\to 0$ with the prescription $1/(\beta\omega)^2\to
V/\lambda^2$, where $\lambda=\sqrt{2\pi\beta}$ is the thermal wavelength.

Let us illustrate this point by computing the partition function for the
repulsive point vortex (\ref{hehe}). In the absence of a harmonic well, the spectrum is
the continuous  free spectrum with for eigenstates Bessel functions with
shifted angular momentum $m\to
m-\alpha$.
It is quite tedious and subtle to compute the partition function
directly in the thermodynamic limit \cite{georgelin}. Consider rather 
the same problem in an
harmonic well. The spectrum is now discrete with a finite degeneracy
\be E_{n,m}=\omega(2n+|m-\alpha|+1)
\ee
with $n\ge 0$. This is nothing but the spectrum of a two-dimensional harmonic oscillator
with shifted angular momentum  $m\to
m-\alpha$. The partition function reads 
\be Z^{\omega}_{\beta}=
{\cosh\beta\omega(\alpha-1/2)\over2\sinh\beta\omega\sinh\beta\omega/2}\ee
Of course it diverges when $\omega\to 0$, but the difference \cite{georgelin}
\be Z^{\omega}_{\beta}-Z^{o\omega}_{\beta}=
{\sinh\beta\omega\alpha/2\sinh\beta\omega(\alpha-1)/2\over\sinh\beta\omega
\sinh\beta\omega/2}\to_{\omega\to 0}{\alpha(\alpha-1)\over 2}\ee 
remains finite\footnote{The same result can be obtained by Zeta function
regularization - see the second reference CMO in \cite{vo}.}
in the thermodynamic limit and yields
(\ref{hehe}).

The partition function
for an
homogeneous  magnetic field (\ref{n16}) can be obtained along the same lines,
with the spectrum
\be\label{ouf} E_{nm}=\omega_t(2n+|m|+1)-\omega_c m\ee
and in the case repulsive point vortex + homogeneous $B$ field,  with the spectrum
\be\label{toto} E_{nm}=\omega_t(2n+|m-\alpha|+1)-\omega_c(m-\alpha)\ee
In the latter case, one  computes 
$Z^{\omega}_{\beta}(B,\alpha)-Z^{\omega}_{\beta}(B,0)$ to obtain,
in the thermodynamic limit $\omega\to 0$, (\ref{n18}).

\subsection{Naive Landau counting rule}

The so-called ``Landau counting rule'' in terms of individual 
states happens
to lead to sometimes uncorrect results, 
although one sums only over a finite number of Landau quantum states.
Again,  correct results are recovered  if the harmonic
regularization is used instead.

Consider for example the spectrum (\ref{toto}) where,
in the thermodynamic limit $\omega\to 0$, i.e. $\omega_t\to
\omega_c$,
the vortex induces a modification of the
Landau spectrum  only for a finite number of states of energy
$\omega_c(2j+1+2\alpha), j\ge 0$, namely the states of orbital angular
momentum  $-j\le m\le0$. Computing the partition function 
(\ref{toto}) directly in the thermodynamic limit with this shifted 
Landau spectrum would lead to
a result different from (\ref{n18}).
The harmonic regulator adds a contribution to the partition function from the states
$m>0$ in (\ref{toto}), that remains finite in the thermodynamic limit, eventhough it is
an infinite sum of vanishing terms, and yields the
correct expression (\ref{n18}).

One can illustrate the failure of the ``Landau counting rule'' on
an other example.
Consider the thermalized orbital persistent current for an homogneous
magnetic field (\ref{n19}) 
\be \label{Imicr}
I^{orb}_{\beta}={1\over Z_{\beta}}
\sum_{n=0}^{\infty}\sum_{m=-\infty}^{\infty}I^{orb}_{n,m} e^{-\beta E_{n,m}}
\ee
in terms of the individual persistent current $I_{n,m}$ for the Landau 
quantum states $\ket{n,m}$ of orbital momentum $m$ and energy  
$E_{n,m}=(2n+1+|m|)\omega_c-m\omega_c$ (i.e. (\ref{ouf}) in the
thermodynamic limit). The individual currents write
\bea
I^{orb}_{n,m}\equiv \frac{e}{2\pi}\bra{n,m}\frac{v_\theta}{r}\ket{n,m}
&=&{e\omega_c\over 2\pi}({m\over|m|}-1)  \hspace{.5cm}\mbox{ if $m\neq0$}\nonumber\\
&=&\frac{e\omega_c}{2\pi}(-1)            \hspace{1.3cm}\mbox{ if $m=0$}
\eea
Despite the fact that the number of states in a given Landau level is
infinite, again only a 
finite number of states  contribute to the persistent current for each
Landau level. Thus the sum (\ref{Imicr}) is convergent. Yet, following this
procedure,  one gets the uncorrect result 
$I^{orb}_{\beta}=-{e\over 2V}\coth b$.
This is again due to the fact that, in this naive ``Landau counting rule'',
if the
$m> 0$ states yield vanishing individual currents, they are still an
infinite number of them. To
give an unambiguous meaning to this infinite summation of vanishing
contributions,  a long distance regulator
is needed, bearing in mind that the thermodynamic limit should be taken 
afterwards the summation over the quantum numbers $n,m$. Indeed, if one 
adds  the harmonic regulator to the Landau Hamiltonian  
(the spectrum is now given by (\ref{ouf})), 
one finds
\bea
I^{orb}_{n,m}&=&{e\omega_t\over 2\pi}({m\over|m|}-{\omega_c\over \omega_t})
\hspace{.5cm}\mbox{ if $m\neq0$}\nonumber\\
&=&\frac{e\omega_c}{2\pi}(-1) 
\hspace{1.6cm}\mbox{ if $m=0$}
\eea
where $\omega_t=\sqrt{\omega^2+\omega_c^2}$. Now the $m>0$ do have
individual non vanishing persistent currents,  and their total contribution
to the thermalized current does not vanish in the thermodynamic
limit. It  is precisely equal to ${e\over 2V}{1\over b}$, a contribution
which,
when added to those of the  $m\le 0$ states, yield
the correct result (\ref{n19}) for $I^{orb}_{\beta}$.

The same situation happens for the thermalized current  induced in the plane
by a repulsive point vortex superposed to an homogeneous $\vec B$ field with
individual currents  
\be I^{orb}_{n,m}={e\omega_t\over 2\pi}({m-\alpha\over|m-\alpha|}-{\omega_t\over\omega_c})\ee

\end{appendix}
\section*{
Acknowledgements}

We would like to acknowledge useful conversations with E.
Akkermans, G. Montambeaux, R. Narevich and D. Ullmo.


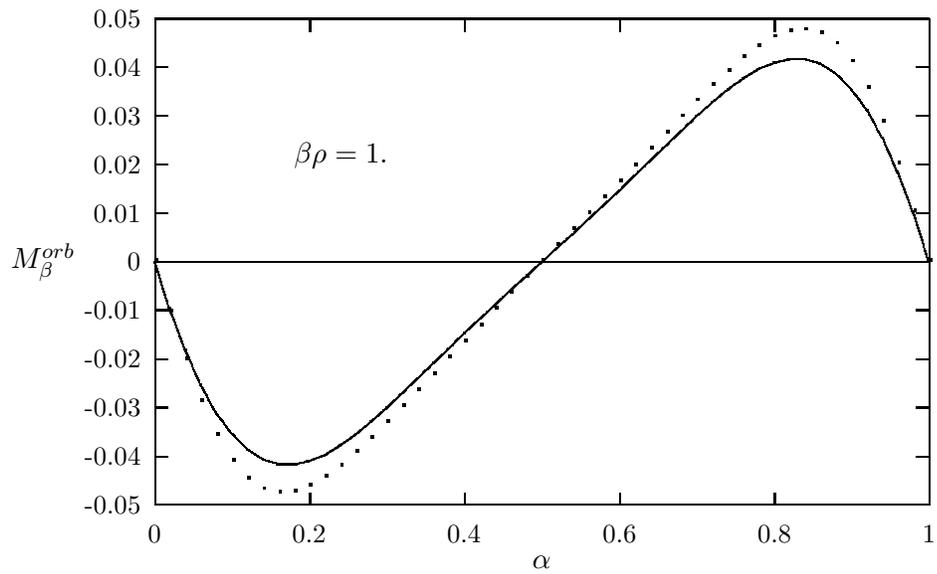
\begin{figure}
\begin{center}


\setlength{\unitlength}{0.240900pt}
\ifx\plotpoint\undefined\newsavebox{\plotpoint}\fi
\sbox{\plotpoint}{\rule[-0.200pt]{0.400pt}{0.400pt}}%
\begin{picture}(1500,900)(0,0)
\font\gnuplot=cmr10 at 10pt
\gnuplot
\sbox{\plotpoint}{\rule[-0.200pt]{0.400pt}{0.400pt}}%
\put(220.0,495.0){\rule[-0.200pt]{292.934pt}{0.400pt}}
\put(220.0,113.0){\rule[-0.200pt]{0.400pt}{184.048pt}}
\put(220.0,113.0){\rule[-0.200pt]{4.818pt}{0.400pt}}
\put(198,113){\makebox(0,0)[r]{-0.05}}
\put(1416.0,113.0){\rule[-0.200pt]{4.818pt}{0.400pt}}
\put(220.0,189.0){\rule[-0.200pt]{4.818pt}{0.400pt}}
\put(198,189){\makebox(0,0)[r]{-0.04}}
\put(1416.0,189.0){\rule[-0.200pt]{4.818pt}{0.400pt}}
\put(220.0,266.0){\rule[-0.200pt]{4.818pt}{0.400pt}}
\put(198,266){\makebox(0,0)[r]{-0.03}}
\put(1416.0,266.0){\rule[-0.200pt]{4.818pt}{0.400pt}}
\put(220.0,342.0){\rule[-0.200pt]{4.818pt}{0.400pt}}
\put(198,342){\makebox(0,0)[r]{-0.02}}
\put(1416.0,342.0){\rule[-0.200pt]{4.818pt}{0.400pt}}
\put(220.0,419.0){\rule[-0.200pt]{4.818pt}{0.400pt}}
\put(198,419){\makebox(0,0)[r]{-0.01}}
\put(1416.0,419.0){\rule[-0.200pt]{4.818pt}{0.400pt}}
\put(220.0,495.0){\rule[-0.200pt]{4.818pt}{0.400pt}}
\put(198,495){\makebox(0,0)[r]{0}}
\put(1416.0,495.0){\rule[-0.200pt]{4.818pt}{0.400pt}}
\put(220.0,571.0){\rule[-0.200pt]{4.818pt}{0.400pt}}
\put(198,571){\makebox(0,0)[r]{0.01}}
\put(1416.0,571.0){\rule[-0.200pt]{4.818pt}{0.400pt}}
\put(220.0,648.0){\rule[-0.200pt]{4.818pt}{0.400pt}}
\put(198,648){\makebox(0,0)[r]{0.02}}
\put(1416.0,648.0){\rule[-0.200pt]{4.818pt}{0.400pt}}
\put(220.0,724.0){\rule[-0.200pt]{4.818pt}{0.400pt}}
\put(198,724){\makebox(0,0)[r]{0.03}}
\put(1416.0,724.0){\rule[-0.200pt]{4.818pt}{0.400pt}}
\put(220.0,801.0){\rule[-0.200pt]{4.818pt}{0.400pt}}
\put(198,801){\makebox(0,0)[r]{0.04}}
\put(1416.0,801.0){\rule[-0.200pt]{4.818pt}{0.400pt}}
\put(220.0,877.0){\rule[-0.200pt]{4.818pt}{0.400pt}}
\put(198,877){\makebox(0,0)[r]{0.05}}
\put(1416.0,877.0){\rule[-0.200pt]{4.818pt}{0.400pt}}
\put(220.0,113.0){\rule[-0.200pt]{0.400pt}{4.818pt}}
\put(220,68){\makebox(0,0){0}}
\put(220.0,857.0){\rule[-0.200pt]{0.400pt}{4.818pt}}
\put(463.0,113.0){\rule[-0.200pt]{0.400pt}{4.818pt}}
\put(463,68){\makebox(0,0){0.2}}
\put(463.0,857.0){\rule[-0.200pt]{0.400pt}{4.818pt}}
\put(706.0,113.0){\rule[-0.200pt]{0.400pt}{4.818pt}}
\put(706,68){\makebox(0,0){0.4}}
\put(706.0,857.0){\rule[-0.200pt]{0.400pt}{4.818pt}}
\put(950.0,113.0){\rule[-0.200pt]{0.400pt}{4.818pt}}
\put(950,68){\makebox(0,0){0.6}}
\put(950.0,857.0){\rule[-0.200pt]{0.400pt}{4.818pt}}
\put(1193.0,113.0){\rule[-0.200pt]{0.400pt}{4.818pt}}
\put(1193,68){\makebox(0,0){0.8}}
\put(1193.0,857.0){\rule[-0.200pt]{0.400pt}{4.818pt}}
\put(1436.0,113.0){\rule[-0.200pt]{0.400pt}{4.818pt}}
\put(1436,68){\makebox(0,0){1}}
\put(1436.0,857.0){\rule[-0.200pt]{0.400pt}{4.818pt}}
\put(220.0,113.0){\rule[-0.200pt]{292.934pt}{0.400pt}}
\put(1436.0,113.0){\rule[-0.200pt]{0.400pt}{184.048pt}}
\put(220.0,877.0){\rule[-0.200pt]{292.934pt}{0.400pt}}
\put(45,495){\makebox(0,0){$M^{orb}_{\beta}$}}
\put(828,23){\makebox(0,0){$\alpha$}}
\put(439,663){\makebox(0,0)[l]{$\beta\rho =1.$}}
\put(220.0,113.0){\rule[-0.200pt]{0.400pt}{184.048pt}}
\put(220,495){\usebox{\plotpoint}}
\multiput(220.58,489.40)(0.496,-1.577){45}{\rule{0.120pt}{1.350pt}}
\multiput(219.17,492.20)(24.000,-72.198){2}{\rule{0.400pt}{0.675pt}}
\multiput(244.58,415.27)(0.497,-1.310){47}{\rule{0.120pt}{1.140pt}}
\multiput(243.17,417.63)(25.000,-62.634){2}{\rule{0.400pt}{0.570pt}}
\multiput(269.58,350.85)(0.496,-1.133){45}{\rule{0.120pt}{1.000pt}}
\multiput(268.17,352.92)(24.000,-51.924){2}{\rule{0.400pt}{0.500pt}}
\multiput(293.58,297.54)(0.496,-0.921){45}{\rule{0.120pt}{0.833pt}}
\multiput(292.17,299.27)(24.000,-42.270){2}{\rule{0.400pt}{0.417pt}}
\multiput(317.58,254.39)(0.497,-0.661){47}{\rule{0.120pt}{0.628pt}}
\multiput(316.17,255.70)(25.000,-31.697){2}{\rule{0.400pt}{0.314pt}}
\multiput(342.00,222.92)(0.498,-0.496){45}{\rule{0.500pt}{0.120pt}}
\multiput(342.00,223.17)(22.962,-24.000){2}{\rule{0.250pt}{0.400pt}}
\multiput(366.00,198.92)(0.753,-0.494){29}{\rule{0.700pt}{0.119pt}}
\multiput(366.00,199.17)(22.547,-16.000){2}{\rule{0.350pt}{0.400pt}}
\multiput(390.00,182.93)(1.865,-0.485){11}{\rule{1.529pt}{0.117pt}}
\multiput(390.00,183.17)(21.827,-7.000){2}{\rule{0.764pt}{0.400pt}}
\multiput(439.00,177.59)(2.602,0.477){7}{\rule{2.020pt}{0.115pt}}
\multiput(439.00,176.17)(19.807,5.000){2}{\rule{1.010pt}{0.400pt}}
\multiput(463.00,182.58)(1.277,0.491){17}{\rule{1.100pt}{0.118pt}}
\multiput(463.00,181.17)(22.717,10.000){2}{\rule{0.550pt}{0.400pt}}
\multiput(488.00,192.58)(0.805,0.494){27}{\rule{0.740pt}{0.119pt}}
\multiput(488.00,191.17)(22.464,15.000){2}{\rule{0.370pt}{0.400pt}}
\multiput(512.00,207.58)(0.708,0.495){31}{\rule{0.665pt}{0.119pt}}
\multiput(512.00,206.17)(22.620,17.000){2}{\rule{0.332pt}{0.400pt}}
\multiput(536.00,224.58)(0.600,0.496){37}{\rule{0.580pt}{0.119pt}}
\multiput(536.00,223.17)(22.796,20.000){2}{\rule{0.290pt}{0.400pt}}
\multiput(560.00,244.58)(0.567,0.496){41}{\rule{0.555pt}{0.120pt}}
\multiput(560.00,243.17)(23.849,22.000){2}{\rule{0.277pt}{0.400pt}}
\multiput(585.00,266.58)(0.520,0.496){43}{\rule{0.517pt}{0.120pt}}
\multiput(585.00,265.17)(22.926,23.000){2}{\rule{0.259pt}{0.400pt}}
\multiput(609.00,289.58)(0.520,0.496){43}{\rule{0.517pt}{0.120pt}}
\multiput(609.00,288.17)(22.926,23.000){2}{\rule{0.259pt}{0.400pt}}
\multiput(633.00,312.58)(0.519,0.496){45}{\rule{0.517pt}{0.120pt}}
\multiput(633.00,311.17)(23.928,24.000){2}{\rule{0.258pt}{0.400pt}}
\multiput(658.00,336.58)(0.520,0.496){43}{\rule{0.517pt}{0.120pt}}
\multiput(658.00,335.17)(22.926,23.000){2}{\rule{0.259pt}{0.400pt}}
\multiput(682.00,359.58)(0.498,0.496){45}{\rule{0.500pt}{0.120pt}}
\multiput(682.00,358.17)(22.962,24.000){2}{\rule{0.250pt}{0.400pt}}
\multiput(706.00,383.58)(0.542,0.496){43}{\rule{0.535pt}{0.120pt}}
\multiput(706.00,382.17)(23.890,23.000){2}{\rule{0.267pt}{0.400pt}}
\multiput(731.00,406.58)(0.544,0.496){41}{\rule{0.536pt}{0.120pt}}
\multiput(731.00,405.17)(22.887,22.000){2}{\rule{0.268pt}{0.400pt}}
\multiput(755.00,428.58)(0.520,0.496){43}{\rule{0.517pt}{0.120pt}}
\multiput(755.00,427.17)(22.926,23.000){2}{\rule{0.259pt}{0.400pt}}
\multiput(779.00,451.58)(0.567,0.496){41}{\rule{0.555pt}{0.120pt}}
\multiput(779.00,450.17)(23.849,22.000){2}{\rule{0.277pt}{0.400pt}}
\multiput(804.00,473.58)(0.544,0.496){41}{\rule{0.536pt}{0.120pt}}
\multiput(804.00,472.17)(22.887,22.000){2}{\rule{0.268pt}{0.400pt}}
\multiput(828.00,495.58)(0.544,0.496){41}{\rule{0.536pt}{0.120pt}}
\multiput(828.00,494.17)(22.887,22.000){2}{\rule{0.268pt}{0.400pt}}
\multiput(852.00,517.58)(0.567,0.496){41}{\rule{0.555pt}{0.120pt}}
\multiput(852.00,516.17)(23.849,22.000){2}{\rule{0.277pt}{0.400pt}}
\multiput(877.00,539.58)(0.520,0.496){43}{\rule{0.517pt}{0.120pt}}
\multiput(877.00,538.17)(22.926,23.000){2}{\rule{0.259pt}{0.400pt}}
\multiput(901.00,562.58)(0.544,0.496){41}{\rule{0.536pt}{0.120pt}}
\multiput(901.00,561.17)(22.887,22.000){2}{\rule{0.268pt}{0.400pt}}
\multiput(925.00,584.58)(0.542,0.496){43}{\rule{0.535pt}{0.120pt}}
\multiput(925.00,583.17)(23.890,23.000){2}{\rule{0.267pt}{0.400pt}}
\multiput(950.00,607.58)(0.498,0.496){45}{\rule{0.500pt}{0.120pt}}
\multiput(950.00,606.17)(22.962,24.000){2}{\rule{0.250pt}{0.400pt}}
\multiput(974.00,631.58)(0.520,0.496){43}{\rule{0.517pt}{0.120pt}}
\multiput(974.00,630.17)(22.926,23.000){2}{\rule{0.259pt}{0.400pt}}
\multiput(998.00,654.58)(0.519,0.496){45}{\rule{0.517pt}{0.120pt}}
\multiput(998.00,653.17)(23.928,24.000){2}{\rule{0.258pt}{0.400pt}}
\multiput(1023.00,678.58)(0.520,0.496){43}{\rule{0.517pt}{0.120pt}}
\multiput(1023.00,677.17)(22.926,23.000){2}{\rule{0.259pt}{0.400pt}}
\multiput(1047.00,701.58)(0.520,0.496){43}{\rule{0.517pt}{0.120pt}}
\multiput(1047.00,700.17)(22.926,23.000){2}{\rule{0.259pt}{0.400pt}}
\multiput(1071.00,724.58)(0.567,0.496){41}{\rule{0.555pt}{0.120pt}}
\multiput(1071.00,723.17)(23.849,22.000){2}{\rule{0.277pt}{0.400pt}}
\multiput(1096.00,746.58)(0.600,0.496){37}{\rule{0.580pt}{0.119pt}}
\multiput(1096.00,745.17)(22.796,20.000){2}{\rule{0.290pt}{0.400pt}}
\multiput(1120.00,766.58)(0.708,0.495){31}{\rule{0.665pt}{0.119pt}}
\multiput(1120.00,765.17)(22.620,17.000){2}{\rule{0.332pt}{0.400pt}}
\multiput(1144.00,783.58)(0.805,0.494){27}{\rule{0.740pt}{0.119pt}}
\multiput(1144.00,782.17)(22.464,15.000){2}{\rule{0.370pt}{0.400pt}}
\multiput(1168.00,798.58)(1.277,0.491){17}{\rule{1.100pt}{0.118pt}}
\multiput(1168.00,797.17)(22.717,10.000){2}{\rule{0.550pt}{0.400pt}}
\multiput(1193.00,808.59)(2.602,0.477){7}{\rule{2.020pt}{0.115pt}}
\multiput(1193.00,807.17)(19.807,5.000){2}{\rule{1.010pt}{0.400pt}}
\put(415.0,177.0){\rule[-0.200pt]{5.782pt}{0.400pt}}
\multiput(1241.00,811.93)(1.865,-0.485){11}{\rule{1.529pt}{0.117pt}}
\multiput(1241.00,812.17)(21.827,-7.000){2}{\rule{0.764pt}{0.400pt}}
\multiput(1266.00,804.92)(0.753,-0.494){29}{\rule{0.700pt}{0.119pt}}
\multiput(1266.00,805.17)(22.547,-16.000){2}{\rule{0.350pt}{0.400pt}}
\multiput(1290.00,788.92)(0.498,-0.496){45}{\rule{0.500pt}{0.120pt}}
\multiput(1290.00,789.17)(22.962,-24.000){2}{\rule{0.250pt}{0.400pt}}
\multiput(1314.58,763.39)(0.497,-0.661){47}{\rule{0.120pt}{0.628pt}}
\multiput(1313.17,764.70)(25.000,-31.697){2}{\rule{0.400pt}{0.314pt}}
\multiput(1339.58,729.54)(0.496,-0.921){45}{\rule{0.120pt}{0.833pt}}
\multiput(1338.17,731.27)(24.000,-42.270){2}{\rule{0.400pt}{0.417pt}}
\multiput(1363.58,684.85)(0.496,-1.133){45}{\rule{0.120pt}{1.000pt}}
\multiput(1362.17,686.92)(24.000,-51.924){2}{\rule{0.400pt}{0.500pt}}
\multiput(1387.58,630.27)(0.497,-1.310){47}{\rule{0.120pt}{1.140pt}}
\multiput(1386.17,632.63)(25.000,-62.634){2}{\rule{0.400pt}{0.570pt}}
\multiput(1412.58,564.40)(0.496,-1.577){45}{\rule{0.120pt}{1.350pt}}
\multiput(1411.17,567.20)(24.000,-72.198){2}{\rule{0.400pt}{0.675pt}}
\put(1217.0,813.0){\rule[-0.200pt]{5.782pt}{0.400pt}}
\put(220,495){\rule{1pt}{1pt}}
\put(244,416){\rule{1pt}{1pt}}
\put(269,341){\rule{1pt}{1pt}}
\put(293,276){\rule{1pt}{1pt}}
\put(317,222){\rule{1pt}{1pt}}
\put(342,181){\rule{1pt}{1pt}}
\put(366,153){\rule{1pt}{1pt}}
\put(390,137){\rule{1pt}{1pt}}
\put(415,131){\rule{1pt}{1pt}}
\put(439,133){\rule{1pt}{1pt}}
\put(463,142){\rule{1pt}{1pt}}
\put(488,157){\rule{1pt}{1pt}}
\put(512,174){\rule{1pt}{1pt}}
\put(536,196){\rule{1pt}{1pt}}
\put(560,218){\rule{1pt}{1pt}}
\put(585,242){\rule{1pt}{1pt}}
\put(609,267){\rule{1pt}{1pt}}
\put(633,293){\rule{1pt}{1pt}}
\put(658,318){\rule{1pt}{1pt}}
\put(682,344){\rule{1pt}{1pt}}
\put(706,369){\rule{1pt}{1pt}}
\put(731,394){\rule{1pt}{1pt}}
\put(755,420){\rule{1pt}{1pt}}
\put(779,445){\rule{1pt}{1pt}}
\put(804,470){\rule{1pt}{1pt}}
\put(828,495){\rule{1pt}{1pt}}
\put(852,520){\rule{1pt}{1pt}}
\put(877,545){\rule{1pt}{1pt}}
\put(901,570){\rule{1pt}{1pt}}
\put(925,596){\rule{1pt}{1pt}}
\put(950,621){\rule{1pt}{1pt}}
\put(974,646){\rule{1pt}{1pt}}
\put(998,672){\rule{1pt}{1pt}}
\put(1023,697){\rule{1pt}{1pt}}
\put(1047,723){\rule{1pt}{1pt}}
\put(1071,748){\rule{1pt}{1pt}}
\put(1096,772){\rule{1pt}{1pt}}
\put(1120,794){\rule{1pt}{1pt}}
\put(1144,816){\rule{1pt}{1pt}}
\put(1168,833){\rule{1pt}{1pt}}
\put(1193,848){\rule{1pt}{1pt}}
\put(1217,857){\rule{1pt}{1pt}}
\put(1241,859){\rule{1pt}{1pt}}
\put(1266,853){\rule{1pt}{1pt}}
\put(1290,837){\rule{1pt}{1pt}}
\put(1314,809){\rule{1pt}{1pt}}
\put(1339,768){\rule{1pt}{1pt}}
\put(1363,714){\rule{1pt}{1pt}}
\put(1387,649){\rule{1pt}{1pt}}
\put(1412,574){\rule{1pt}{1pt}}
\put(1436,495){\rule{1pt}{1pt}}
\end{picture}


\caption{The orbital magnetization in unit $e=1$ 
in  the magnetic impurity problem 
for $\beta\rho =1$. Comparison between the analytical   computation 
(\ref{n34}), full curve,  and numerical   simulations (\ref{n31}). 
The points were obtained by generating 2000 curves of 100000 steps each one. 
}

\end{center}
\end{figure}

\end{document}